\documentstyle[11pt,newpasp,twoside]{article}
\def\edcomment#1{\iffalse\marginpar{\raggedright\sl#1\/}\else\relax\fi}
\reversemarginpar

\begin{document}
\title{Stellar wind signatures in sdB stars?}
 \author{U. Heber}
\affil{Dr. Remeis-Sternwarte, Astronomisches Institut, Universit\"at 
Erlangen-N\"urnberg, Bamberg, Germany}
\author{P.F.L. Maxted}
\affil{School of Chemistry \& Physics, Keele University, GB}
\author{T.R. Marsh, C. Knigge}
\affil{University of Southampton, Department of Physics \& Astronomy, GB}
\author{J. E. Drew}
\affil{Imperial College of Science, Technology and Medicine, London, GB}

\begin{abstract}
Subdwarf B (sdB) stars form the blue end of the horizonal branch. Their 
peculiar atmospheric abundance patterns are due to diffusion processes.
However, diffusion models fail to explain these anomalies quantitatively.
From a NLTE model atmosphere analysis of 40 sdB stars,   
we found that the more luminous (i.e. more evolved) stars have 
anomalous H$\alpha$ and HeI\,6678\AA\ line profiles, i.e, the lines are too
broad and shallow and may even show some emission. We interpret these
anomalies as the signatures of a stellar wind, the first such detection in
this class of star (if confirmed).  Mass loss
may also explain the peculiar abundance patterns seen in sdB stars.
High-quality UV spectra are needed to confirm that these stars do have 
stellar winds. 
\end{abstract}

\section{Introduction}

Subdwarf B (sdB) stars are core helium burning stars with very thin hydrogen
envelopes which all have masses of 0.46-0.50 solar masses and can be 
identified with models of extreme horizontal Branch (EHB) stars (Heber, 
1986). 

\begin{figure}
\vspace{8cm}
\includegraphics{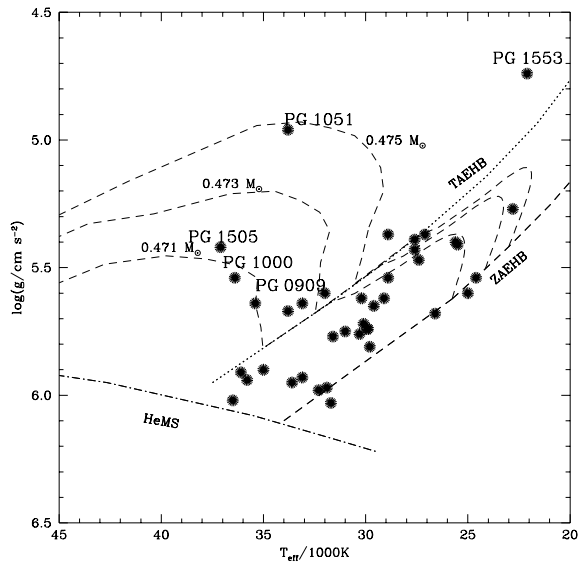}
\caption{(T$_{\rm eff}$,log\,g) diagram for the Maxted et al. (2001) sdB sample. 
Note that most stars are well matched by models for the extreme horizontal 
branch. Five stars (labelled by their first 4 digits of their PG catalog 
names) have already evolved from the EHB.}\label{hrd}
\end{figure} 

The atmospheres of sdB stars are typically deficient in helium by more than one 
order of magnitude, but 
there is a wide spread from solar He/H to stars
with He/H$<$10$^{-4}$. These He abundances are still much too large to be 
accounted for by diffusion, i.e. by the balance between radiative levitation 
and gravity (Michaud et al., 1989)
which predicts He abundances {\bf lower} by
two orders of magnitudes than the average observed He abundance 
of He/H$\approx$10$^{-2}$ (see Fontaine \& Chayer, 1997).
Since the diffusion timescale is small (10$^4$ yrs) compared to the EHB 
lifetime (10$^8$ yrs), {\bf no} He should be visible in 
the radiative atmospheres: the 
puzzle here is not so much a lack of He but an excess 
in the sdB atmospheres. 

Modelling of metal diffusion in sdB stars is still in 
its infancy. First attempts have been published by Charpinet et al. (1997)
for Fe and Unglaub \& Bues (2001) for C, N and O but met the observational 
constraints with very little success (except for iron). 

\section{Diffusion and mass loss}

Stellar winds have
frequently been suggested as an explanation.
The first realistic calculations 
have been carried out by Fontaine \& Chayer (1997) and Unglaub \& Bues 
(1998, 2001).
The observed He abundances can indeed be explained 
if a mass loss rate of 10$^{-12}$ to 10$^{-14}$ 
Msolar/yr is adopted. 

Yet it is not clear whether diffusion models incorporating mass 
loss can explain the metal abundance anomalies, at the same time as the He 
abundance.

Mass loss has been detected in sdO stars which are, however, considerably 
more luminous than sdB stars. HD128220B (T$_{\rm eff}$=40600K, log\,g=4.5,
Rauch, 1993) for example displays strong, broad, blue shifted N~V 
lines produced in its stellar wind. 

Since sdB stars are less luminous we expect the mass loss rates to be lower 
than in sdO stars and therefore harder to detect. Indeed, up to now there 
is no observational proof for mass loss and, therefore, the mass loss rate 
is still a free 
parameter for diffusion models. 

First hints about stellar winds came from the 
quantitative analysis of optical blue spectra obtained during the
survey of Maxted et al. (2001). They derived atmospheric parameters
of 40 sdB stars from Balmer (H$\beta$ to H9) and He I lines using 
static NLTE model atmospheres (see Napiwotzki, 1997). 
The results are reproduced in Figure 1.
As can be seen all but five stars fall onto the extreme Horizontal Branch
as expected. The five stars above the EHB in Fig.~1 have probably 
already evolved from the EHB. A comparison of theoretical H$\alpha$ line 
profiles 
(synthesised for the atmospheric parameters derived from the blue spectra, 
see Maxted et al., 2001) 
to the observations revealed perfect matches for all stars except for the
four post-EHB stars (see Fig.~2).  

\begin{figure}
\vspace{10.0cm}
\includegraphics{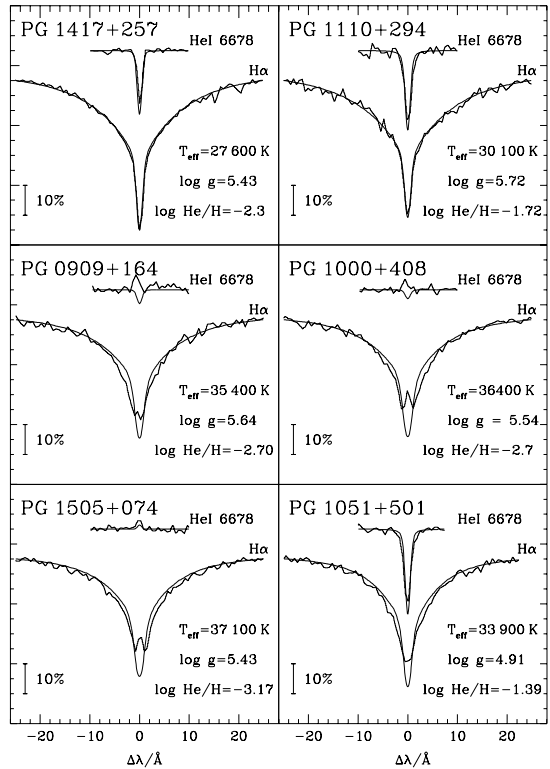}
\caption{Comparison of synthetic line profiles of H$\alpha$ and 
He I, 6678\AA\ for two sdB stars on the EHB (top panels) and the post-EHB
stars PG0909+164, PG1000+408, PG1505+074 and PG1051+501. Note that these 
are not fits to the observed line profiles. Model parameters have been 
derived form the fit of blue spectra. Note also that the H$\alpha$ profiles
in PG0909+164, PG1000+408 (middle panel) are slightly asymmetric.}
\label{halpha}
\end{figure}

By plotting the goodness of fit versus luminosity 
(Fig.~3) of the
stars a clear trend becomes obvious. The quality of the fit deteriorates 
with increasing luminosity. In the optical spectral range H$\alpha$ is the
line most sensitive to stellar winds. Line asymmetries are another 
signature indicative of a stellar wind.

\begin{figure}
\vspace{7.0cm}
\includegraphics{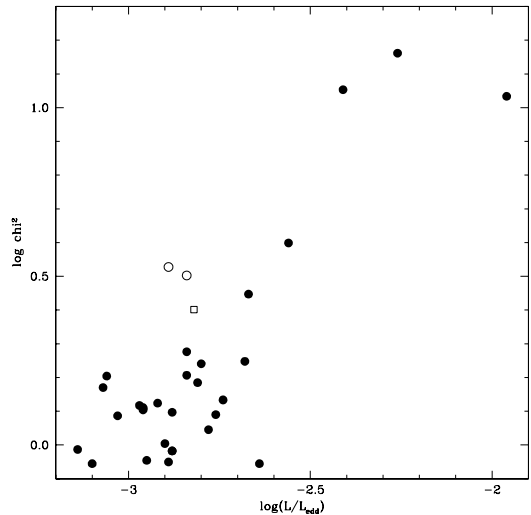}
\caption{Quality of fit (reduced $\chi^2$) as a function of stellar 
luminosity (expressed in units of the Eddington luminosity). The open 
symbols mark known (square) or suspected rotating stars. Rotational 
broadening has not been taken into account.}
\label{luminosity}
\end{figure}

Since mass loss rates of hot stars are expected
to increase with luminosity (Pauldrach et al., 1998), we conjecture 
that the observed 
deviations of the H$\alpha$ lines from the predictions of static NLTE models
for the most luminous stars in our sample
are caused by stellar winds. The somewhat asymmetric profiles of PG0909+164 
and PG1000+408 make these two targets the best candidates.

SdB stars form a fairly homogenous group of stars which evolve to higher 
luminosities (from 10$^{-3.4}$ to 10$^{-2}$ Eddington luminosities, 
see Fig.~3), into the regime where winds switch on. If we 
can establish that some sdB stars do have winds, it will provide an 
important test for models of line-driven winds.  

High-quality UV spectra are needed to establish 
whether these stars really do have stellar winds. 

\end{document}